\documentclass[%
 aip,
 jmp,%
 amsmath,amssymb,
 reprint,%
]{revtex4-1}
\usepackage{graphicx}
\usepackage{dcolumn}
\usepackage{bm}
\usepackage{upgreek} 

\begin{document}


\title[Development of a Virtual EM Detector for the Advanced Particle Accelerator Modeling Code WarpX]{Development of a Virtual EM Detector for the Advanced Particle Accelerator Modeling Code WarpX}

\author{E. Rheaume}
\affiliation{ 
California State University, Long Beach
}%
\affiliation{ 
Lawrence Berkeley National Laboratory
}%
\author{L. Giacomel}
\affiliation{ 
CERN (European Organization for Nuclear Research)
}%
\author{J. Vay}
\author{A. Huebl}
%
\affiliation{ 
Lawrence Berkeley National Laboratory
}%

\date{\today}

\begin{abstract}
In physics research particle accelerators are highly valued, and extraordinarily expensive, technical instruments. The high cost of particle accelerators results from the immense lengths required to accelerate particles to high energies, using radio frequency cavities. A current promising field of research, laser-driven particle acceleration has the potential to reduce the size, cost, and energy consumption of particle accelerators by orders of magnitude. 
To understand and control particle acceleration in plasmas using ultra-small spatial configurations, researchers have been developing computational models to simulate the acceleration environment. Within these models, computational scientists have introduced virtual diagnostics to serve as the digital parallel to experimental detectors.
Using WarpX\cite{Vay:2021}, an advanced Particle-in-Cell code that simulates laser-driven particle acceleration, we have developed a virtual diagnostic to measure electromagnetic radiation. Such radiation can for instance be produced from scattered and transmitted laser beams. This \textit{FieldProbe} diagnostic measures individual field components for electric and magnetic fields at a point or at all points along a specified geometry (line or plane). This diagnostic is GPU-accelerated and parallelized using the Message Passing Interface (MPI) and can thus run on High Performance Computing Centers such as NERSC.

\end{abstract}

\keywords{WarpX, Laser-Ion Acceleration, in-situ diagnostics, Particle-in-Cell, Field Probe}
\maketitle

\section{\label{sec:level1}Introduction}

Particle accelerators are essential tools in industry and natural sciences, from applications in the semiconductor industry and medicine to material science and fundamental research in physics. What determines the size and ultimately the cost of an accelerator are the energies and quality of the particle beam that need to be produced. To achieve energies large enough to study fundamental processes in high-energy physics, today’s particle accelerators are one of the longest, largest, and often most expensive investments constructed by any laboratory. But equally important are small and mid-size accelerators, as used for industrial and medical applications such as imaging, sterilizing food, and irradiating tumor cells.
\begin{figure}
\includegraphics[scale=.4]{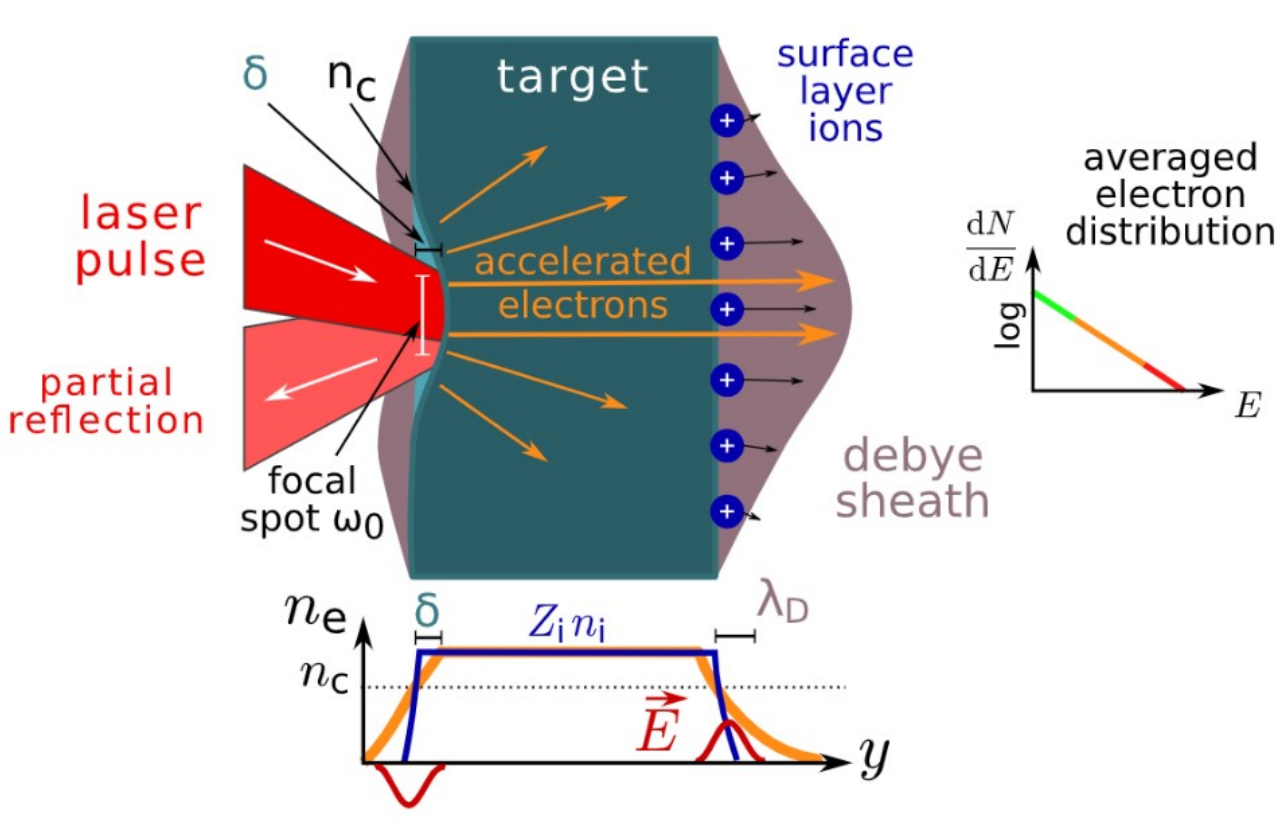}
\caption{\label{fig:wide}Overview of laser absorption into electrons and ion acceleration via target-normal sheet acceleration of ions \cite{Wilks1992,Mora2005}. Figure adopted from Ref.~\cite{huebl:2019}
}
\label{fig:laser-abs}
\end{figure}

While many conventional particle accelerators rely on radio frequency to accelerate particles, their accelerating fields are limited to 100’s of Mega Volts per Meter due to material limitations. Because of this limitation, to-be accelerated particles need to pass through those fields over large distances and, consequently, these accelerators span many miles (SLAC LCLS II: 2 mi; CERN LHC: 16.6 mi; CERN CLIC proposed 11 - 50 km). Due to the high cost and technical challenges of constructing these accelerators, new technology is being researched to substantially decrease the size of future machines. One promising research field is laser-driven ion acceleration that uses ultra-short ($\sim30$\,fs), ultra-intense (PW) laser pulses, which interact with a plasma, separating electrons from their ions to form a pseudo-capacitor, creating an internal acceleration field that is orders of magnitude higher (up to Tera-Volts per Meter) than achievable in existing accelerator technology ($\sim100$\,MV/m). This results in accelerators that span millimeters as opposed to miles.

\begin{figure}
\includegraphics[scale=.45]{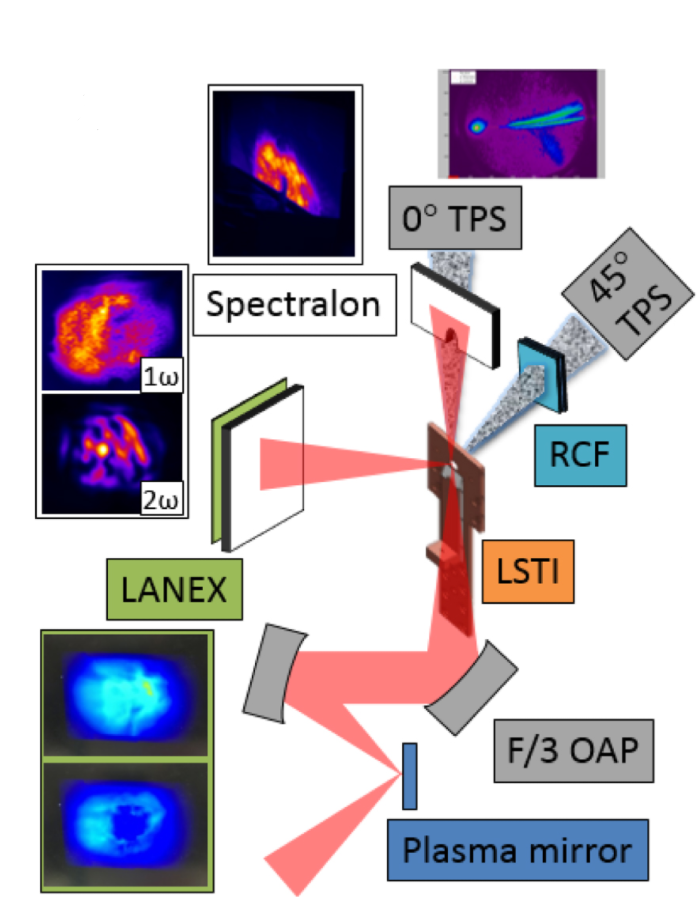}
\caption{Setup of a Laser-Ion Acceleration experiment from Ref.~\cite{Poole_2018}. LSTI (linear slide target inserter) is an aparatus used to swap out target materials. TPS (Thomson parabola spectrometer) devices are used to record accelerated particles. Spectralons are placed around the simulation space to measure reflected and transmitted light. A LANEX sheet is used to record energized electrons, displaying the pattern created by ejected electrons. RCF (radiochromic film) packs are used to record spatial information of accelerated protons. OAP (off-axis parabolic) is the type of mirror used here.}
\label{fig:exp-setup}
\end{figure}

\begin{figure}
\includegraphics[scale=.32]{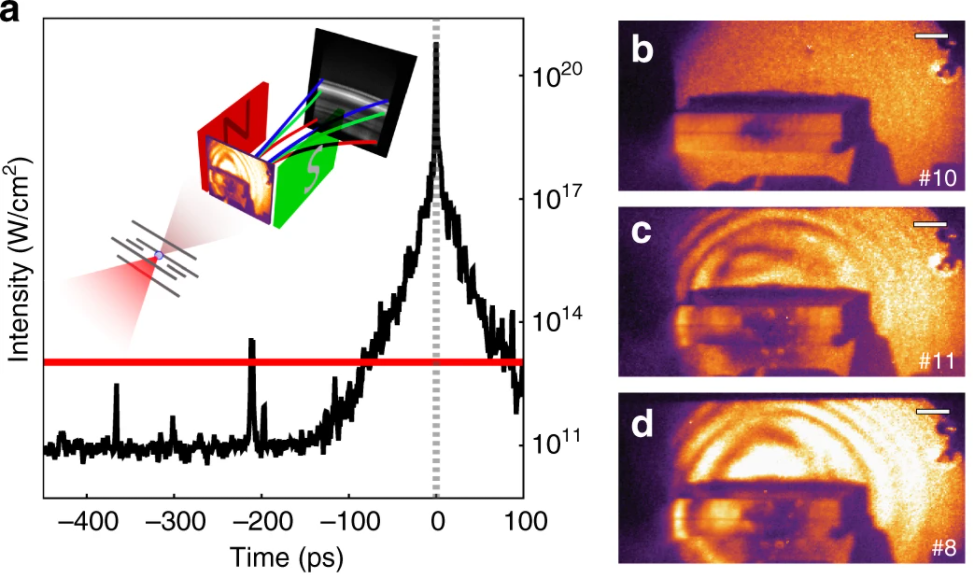}
\caption{Experimental detector panel in Ref.~\cite{Hilz_2018}. The circular diffraction pattern pictured on the right is a common result seen using optical detectors.
}
\label{fig:detect_pan}
\end{figure}

Figure~\ref{fig:laser-abs} shows a schema of laser-driven ion acceleration in the Target-Normal Sheet Acceleration (TNSA) regime~\cite{Wilks1992,Mora2005}.
As the laser pulse (red) interacts with the front surface of an opaque \textit{target}, the transmitted energy ionizes the material to create a plasma. Inside a plasma electrons flow freely dissociated from their ions. While the frequency of the laser pulse is less than or equal to that of the electrons in the plasma, energy from the laser pulse is passed to plasma electrons, pushing them into the opaque region of the target. From the initial kick of the laser-pulse, these electrons propagate to the rear side of the material, creating a scenario where a sheet of net negative charges, with the thickness of the plasma Debye length, covers the rear surface. This creates an electric field similar to that of a capacitor \cite{Wilks1992,Mora2005}. Due to the strength of the generated electric field in a laser-ion accelerator, laser-plasma acceleration processes are ultra-fast (femto- to picosecond scale) and very small (micrometer scale). Because of these fine scales, accurate control over these processes is crucial. To improve these accelerators, computational models are used to accurately simulate the laser-plasma interaction dynamics. In laboratory settings, detector plates can be employed as a diagnostic to measure transmitted and reflected electromagnetic (EM) radiation. Figure~\ref{fig:exp-setup} shows an experimental, laboratory setup of a laser-ion acceleration. In this scenario, the Spectralons as well as the LANEX sheet detect scattered light via electromagnetic radiation. The resulting data is expected to look like figure~\ref{fig:detect_pan} (b-d). In the lab, these optical detectors are important to develop insight into the ultrafast processes happening during the laser-plasma interaction. In a simulated environment, a virtual diagnostics fills this role of measuring physical observables corresponding to critical data from physical accelerators. By joining these computational results with experimental data, researchers can refine the technology behind laser-ion acceleration, resulting in increased viability for this acceleration method.
\section{Materials and Methods}

This work extends WarpX\cite{Vay:2021}, an advanced Particle-In-Cell (PIC) code that simulates laser-plasma interactions using a variety of numerical methods. WarpX in turn utilizes AMREX\cite{zhang:2019}, a software library designed to support the programming of High Performance Computers by managing parallel communication, domain composition, load balancing, mesh data structures, geometric solvers for parabolic systems, and CPU and GPU execution. Both projects are written in modern C++ and are functional on everything from home computers to supercomputers. WarpX uses an open source development model that allows the code to be easily accessed, reviewed, tested and modified to suit the needs of individual scientists.

\begin{figure}
\includegraphics[scale=.8]{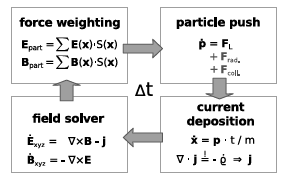}
\caption{The explicit time loop of the electromagnetic particle-in-cell algorithm. Figure from Ref.~\cite{huebl:2019}.
}
\label{fig:EM_PIC}
\end{figure}
\begin{figure}
\includegraphics[scale=.55]{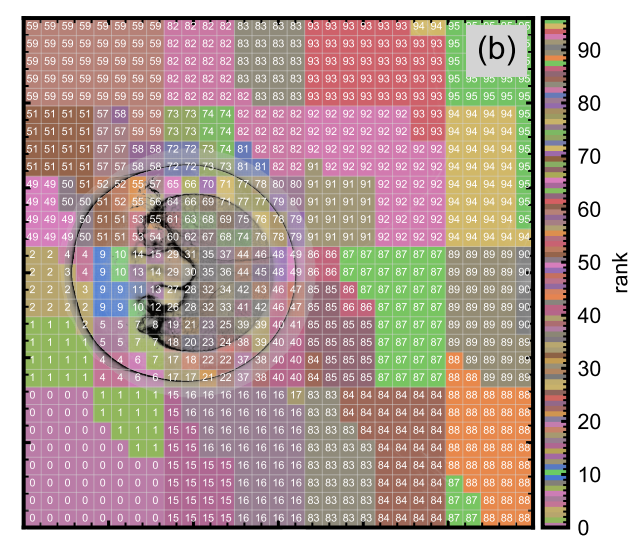}
\caption{Domain decomposition. In WarpX, typically 1 rank is assigned to each GPU. The simulation domain is divided amongst GPUs, allotting more GPUs to regions with more processes. Large regions with little computational requirements can be handled by a single GPU.
Figure from Ref.~\cite{Rowan_2021}.
}
\label{fig:domain_dec}
\end{figure}

The PIC method is an explicit forward iterating scheme that loops in small time steps over a given time frame. The iteration continues until the acceleration process of a few picoseconds has elapsed and all particle energies are converged. The virtual diagnostic is implemented directly in WarpX, thus creating detector results alongside the running simulation (in situ). In situ algorithms are advantageous over conventional output and post-processing workflows because they do not need to save detailed simulation data as the algorithm is running. This results in significantly smaller data output requirements. Once the program has terminated, the final detector result is readily available for immediate visualization and analysis by the scientist.

Because Particle-in-Cell simulations are large, multi-dimensional, and resource intensive programs, it often necessitates the use of parallel computing, the process by which a computationally heavy program's tasks are divided among multiple GPUs to improve the run time. Figure~\ref{fig:domain_dec} illustrates an example of this division, also known as \textit{domain decomposition}. In this figure, the simulation space is pictured broken up cell-by-cell into 96 different ranks. In WarpX, each rank is typically processed by a single GPU. The simulation regions with many particles are allotted more ranks, whereas large regions with few particles are managed by individual ranks. This is the basis of High Performance Computing as utilized in supercomputers.
\begin{figure}
\includegraphics[scale=.18]{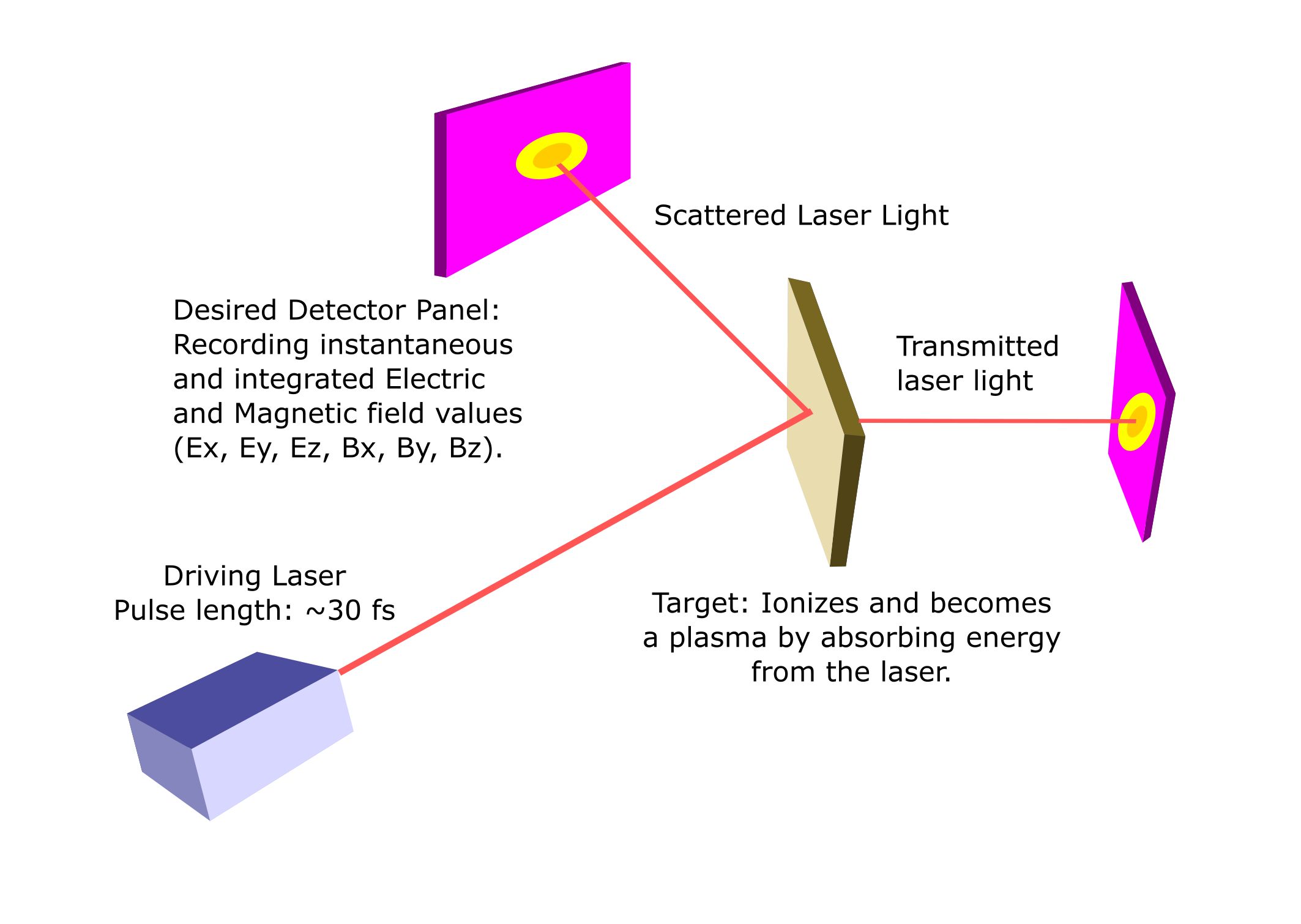}
\caption{Detector target implementation. In this diagram, the detector is positioned behind the plasma to gather reflected light from the laser pulse.
}
\label{fig:desired_det}
\end{figure}
The diagnostic being developed here is a FieldProbe, an in situ detector that measures electric and magnetic field components. Figure~\ref{fig:desired_det} illustrates the development goal: to implement up to 2 dimensional detector plates within the domain of a simulation, able to extract data instantaneously without interfering with the simulation at large. These plates are designed to quantify reflected and transmitted light resulting from the driving laser interacting with the plasma target. In order to use the new diagnostics, WarpX users are presented with additional input options to add one or more virtual diagnostics to their simulation.

\begin{figure*}
\includegraphics[scale=.2]{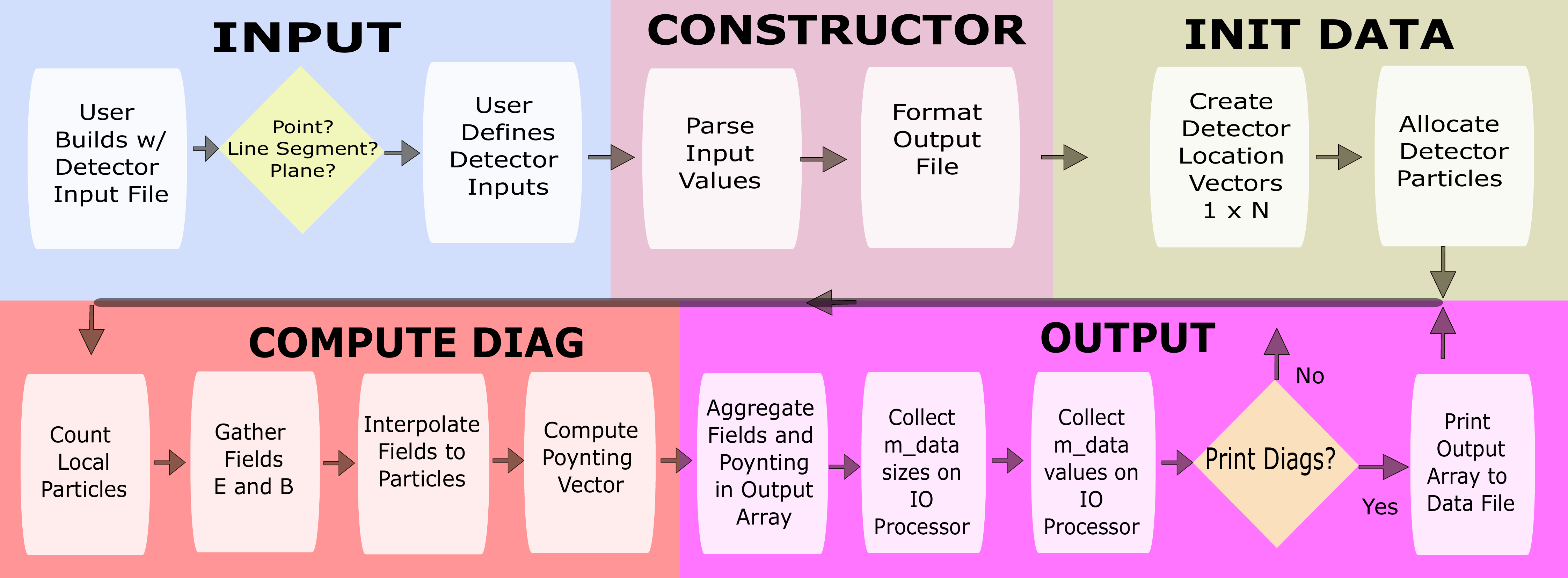}
\caption{Field Probe implementation logic.}
\label{fig:flowchart}
\end{figure*}

The user controls properties of the detector(s) with according parameters, e.g., the actively recording components of the detector can either be a chosen point, a specific line segment, or a plane. (See FIG.~\ref{fig:flowchart} Input and Constructor steps.) In the event of a singular point detector, the user defines the spatial coordinates of the point. At that point, a detector particle is placed in the simulation. In the event of a line segment detector, the user defines the coordinates of a start point, an endpoint, and a resolution factor. The line segment connecting these points is then divided into equally spaced coordinates. At each of these locations a detector particle is placed. In the event of a plane detector (which is only usable during a 3D simulation), the user defines a point, a vector normal to the desired plane, a vector pointing towards the ``top'' of the plane square (to define the square's rotation), the geometric radius of the detector, and a resolution factor. Location data is interpolated from these inputs to create a square of equally spaced particles. Determining the location of detector particles and adding them to the simulation occurs during the Initialize Data step.
Regardless of the detector geometry, each point is handled in a similar fashion. During each iteration of the PIC loop, for each particle, the electric and magnetic fields are fetched. These values are stored temporarily before being used to compute the Poynting vector (EQ: 1), a measure of the electromagnetic flux calculated via the electric field $\vec E$, the magnetic field $\vec B$, and the vacuum permittivity $\mu_{0}$. After the Poynting vector is obtained, all calculated values are stored inside the particle (which can store Struct-of-Array (SoA) and Array-of-Struct (AoS) data).\cite{zhang:2019}
The latter is shown in the Compute Diagnostic step of FIG.~\ref{fig:flowchart}.

During our test case, we set the diagnostic to be performed every 200 steps, during which the observed electric field components, magnetic field components, Poynting vector magnitude, and location data are written to a file for later analysis. If the user prefers, they can set the diagnostic instead to integrate those values over the duration of the simulation. In this case, the diagnostic is run on every simulation step during which the individual EM components are added to the components from previous iterations, effectively summing the electric and magnetic field components over the duration of the experiment. The size of the Poynting vector is aggregated in the same fashion. Every 200 steps, the summed data is multiplied by the elapsed time and output. The data is arranged in a 12 x N — N is the number of detector particles — vector of 8-bit doubles (components simulation time, time step, $r_x$ , $r_y$ , $r_z$ , $E_x$ , $E_y$ , $E_z$ , $B_x$ , $B_y$ , $B_z$ and $|S|$), which is printed into a corresponding data file. These values (left to right) describe the time, position, the summation of $\vec E$ field components, the summation of $\vec B$ field components, and the Poynting vector magnitude for each detector particle. These data files can be analyzed using Jupyter lab (or other python tools) to visualize the overall electric field, magnetic field, and EM wave amplitude.
\begin{equation}
    \label{eq:poynt}
    \vec{S} = \frac{1}{\mu{\scriptscriptstyle 0}}(\vec{E} \times \vec{B})
\end{equation}
\[\mathrm{Poynting~Vector~Calculation}\]

\section{Results (preliminary / expected)}
\subsection{Implementation Details}
The original FieldProbe was designed to parse data required for a single point in the simulation, gather instantaneous values for either raw or interpolated fields at that point, and output the corresponding data. To implement an integration tool as well as to permit multi-dimensional scalability, a particle container is initialized. Each particle in the container stores a combination of Struct-of-Array and Array-of-Struct data which can be passed to functions throughout the program. The particle container is modeled after the \verb+WarpXParticleContainer+, however unused attributes such as particle velocity or weighting were removed and replaced with storage for the electric field, magnetic field, and Poynting vector. When the program first initializes data, the parsed geometric location of the FieldProbe is passed into a function which creates the particles within the container and distributes them to their appropriate CPUs.

When the FieldProbe diagnostic is called, it iterates over all levels (for all current cases, all particles are initialized on level 0) and all particles within each level using a particle iterator. To do this, an interface extension was made to the preexisting \verb+ReducedDiags+ class from which FieldProbe takes its structure. This interface adds a virtual function to initialize data which is called before the diagnostic is computed. This is critical because the particles cannot be initialized during the constructor as the simulation environment is not fully rendered by this point. For each level, the electric and magnetic fields are pulled from the simulation. Then, for each particle, these fields are interpolated onto the particle's position during a \verb+ParallelFor+ loop. This central part is GPU-accelerated for Nvidia, AMD and Intel GPUs. The magnitude of the Poynting vector is calculated, then all 7 calculated values are stored directly into the particles Struct-of-Array data. This data is not lost upon future iterations of the WarpX simulation, allowing for these values to be integrated upon during each time step. For the user-provided interval, e.g., every 200 time steps, the particle data is copied to an output vector.

Node-to-node communication is needed for supercomputers with multiple computing nodes.
The diagnostics uses the  Message Passage Interface (MPI) API for this aspect, when the data vector needs to be written out to a data file for future analysis. Using the AMReX library's built-in \verb+MPI_Gather+ equivalent, each active node sends a single integer to the Input-Output (IO) Processor node. This integer signifies the size of each node's local data vector, the size of which will most likely vary between each node. The IO node uses this information to allot size to a final collective output vector where data will be collected. It also computes where in the vector each parallel node’s data will be written. Once this has been calculated, all nodes initiate an \verb+MPI_Gatherv+ to converge output data onto the IO processor node. This data is then passed to a \verb+WriteToFile+ function which prints the data into an accessible CSV text file. For a simplified analysis, we used Jupyter Lab\cite{Kluyver2016jupyter} to create a Conda analysis environment where we can read, sort, and plot data from this text file. 

The critical design considerations that had to be made regarded the size of the output file as well as the efficiency of the diagnostic. The original FieldProbe exported 6 doubles every 200 steps over a 2000 step simulation. Including the 0th step, this implementation outputs 528 bytes for the entire run. When the integrated diagnostic is employed, the same output method is used. While the diagnostic needed to be run at every step instead of every 200 steps, the data was only written out with the same frequency as the original FieldProbe. By adding location data, as well as the Poynting vector magnitude, the output was increased to 880\,Bytes. Initially outputting the data at every time step was proposed, but that implementation would produce 160\,kB of data. While this is reasonable for one dimensional data, when more particles are introduced, the output data size would balloon exponentially (100s of MB for line detector, 100s of GB for plane detector). Thus, the output was limited to once every 200 steps which, for an image resolution of 1'000 points, creates line data to the order of 880\,kB and plane data to the order of 880,MB. Even so, should the user prefer, the output frequency can be adjusted as an input parameter to allot for heavier or lighter data quantities.

In future updates, the text based output will be replaced with scalable, portable binary output using the openPMD~\cite{openPMD} format.
WarpX already provides interfaces to write openPMD and re-using this functionality will allow to store high-fidelity FieldProbe ensembles.

To maximize the efficiency of the diagnostic, best practices were used whenever available. Every operation involving resizing, filling, communicating, and printing the data vector is set to only run during the time steps that output data. For diagnostic settings that do not integrate and instead return instantaneous values, the diagnostic is not run at all during the intermediary steps. MPI communication involves a larger latency compared to local access of data. To avoid slowing down the diagnostics, final output data is gathered only to the IO processor node, as opposed to sharing the data to all nodes. Operations on the data, as well as temporary vectors used to facilitate communication, exist only on the IO node, thus not consuming resources on assistant nodes.
A downside of this approach is, that the maximum number of probe points is limited to what the IO processor node can receive, with respect to available host memory.
In practice, this is currently no problem as long as one only records probes of a few billion points in total.
Nonetheless, this limitation will easily be lifted when needed, by using aforementioned openPMD output, which writes directly and in parallel from each compute rank without involving an MPI Gather of the written data.

To test the probe, basic input parameters were added to the Reduced Diagnostics test example. This not only checks that the code compiles and produces viable data, but it provides a baseline data set against which future iterations of the code can be compared to ensure that any changes do not disrupt the core function of the code.

\subsection{Evaluation}
The FieldProbe diagnostic has been successfully implemented in 1D, 2D, 3D and RZ (cylindrical) geometry of WarpX. Below the concept is demonstrated in a simulation of the classic 2 dimensional double-slit experiment. Because WarpX already contains functionality for laser pulse initialization, the double slit observables are calculable via a plane electromagnetic wave, and the double slit experiment has been solved analytically, this experiment can serve as a test case for the functionality of the FieldProbe diagnostic. This experiment was developed to prove the quantum phenomena of light that causes it to behave as both a wave and a collection of particles. During the double slit experiment, light (in the form of a plane wave of a specific, known wavelength) passes through a plate with two thin slits (of a specific, known size and separation). As the wave moves through the slits, 2 identical waves, sharing the same offset and wavelength as the plane wave, emanate from each slit. The resulting waves interact constructively at all locations where the distance traveled from the slits respectively are integer multiples of the wavelength. They interact destructively at locations where those distances are integer + $1/2$ multiples of the wavelength. These distances vary with slit separation and wavelength. Beyond the slit plate, a detector plate lies a specific distance from the slit. Because of the constructive and destructive interactions between the light waves originating from each slit, an oscillation of bright (constructive) and dark (destructive) bands appear on the detector plate. The intensity of these spots correlates to the magnitude of the Poynting vector for each of these points.

\begin{figure}
\includegraphics[scale=.6]{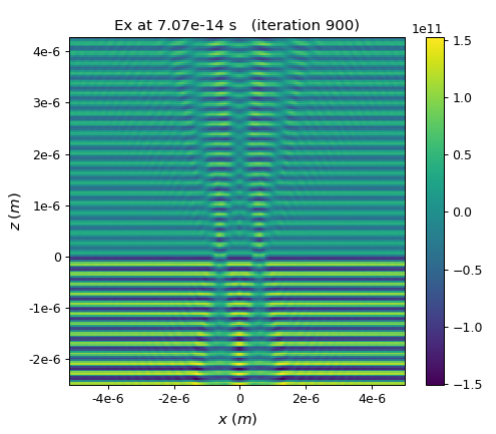}
\caption{The openPMD\cite{openPMD} diagnostic allow researchers to graph particles and fields while the simulation is running.
}
\label{fig:openpmd_double}
\end{figure}
\begin{figure}
\includegraphics[scale=.6]{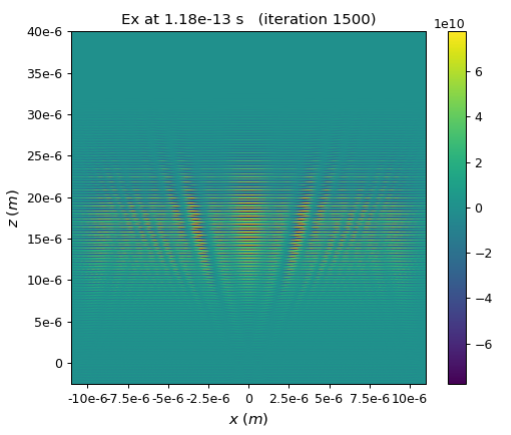}
\caption{The $\vec E$ field can be seen experiencing diffraction as a result of the double slit. A smaller degree of the plane wave permeates the barrier, resulting in a positive EM-Flux offset being recorded.
}
\label{fig:openpmd_waves}
\end{figure}
To produce this test case, we compiled a 2 dimensional version of WarpX. While a 3D test could be used instead, running a 3D simulation requires substantially more processing power, and a 2D test is sufficient to prove the correctness of the FieldProbe. The test geometry is a 10 $\mu$m (x direction) by 20 $\mu$m (z direction) square. A laser pulse of $\lambda = 0.2 \mu$m, modified to resemble a plane wave by defining a large waist distance, is initialized 0.2 $\mu$m before the double slit. The double slit was originally simulated by using a dense particle slab of electrons. Because an electromagnetic PIC loop does not reinitialize point charges (see FIG.~\ref{fig:EM_PIC}), the electron particle density assumes a neutral charge, implying the existence of an infinitely heavy proton background at their initial positions. As the plane wave interacts with this density, the electrons oscillate in time with the wave frequency to prevent the wave from permeating beyond the skin depth of our plasma. This interaction can be seen in FIG.~\ref{fig:openpmd_double}. This causes a double slit effect to occur beyond the particle density. The propagating waves can be seen in FIG.~\ref{fig:openpmd_waves}. This particle density has a critical density of $2.78e28\,\mathrm{m}^{-3}$ and a radius of $15$\,nm. Slits are separated (center to center) by $1.2\,\upmu$m (calculated $\lambda \cdot 6$) and have a width of $0.1\,\upmu$m. The FieldProbe line out instantaneous detector is placed $10\,\upmu$m from the double slit while the integrated line detector is placed $19.5\,\upmu$m.

\begin{figure}
\includegraphics[scale=.7]{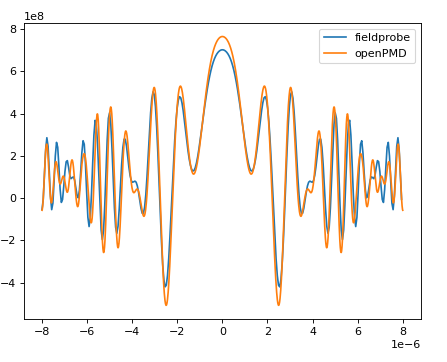}
\caption{Instantaneous line-out Ex field data captured from Field Probe and openPMD at 6.01e-14 seconds.
}
\label{fig:line_out}
\end{figure}

In FIG.~\ref{fig:line_out}, we can observe the efficacy of the FieldProbe instantaneous line out by capturing Ex values at a specified time step and comparing them to openPMD\cite{openPMD}, an existing diagnostic that captures and plots particle and field data. Because the FieldProbe interpolated field data to a physical particle (with a set geometry), the values recorded are sometimes offset by a small margin; in this case, the adjustment is $\lambda/10$. This figure demonstrates that the FieldProbe instantaneous detection is returning values in line with expected results.

\begin{figure}
\includegraphics[scale=.36]{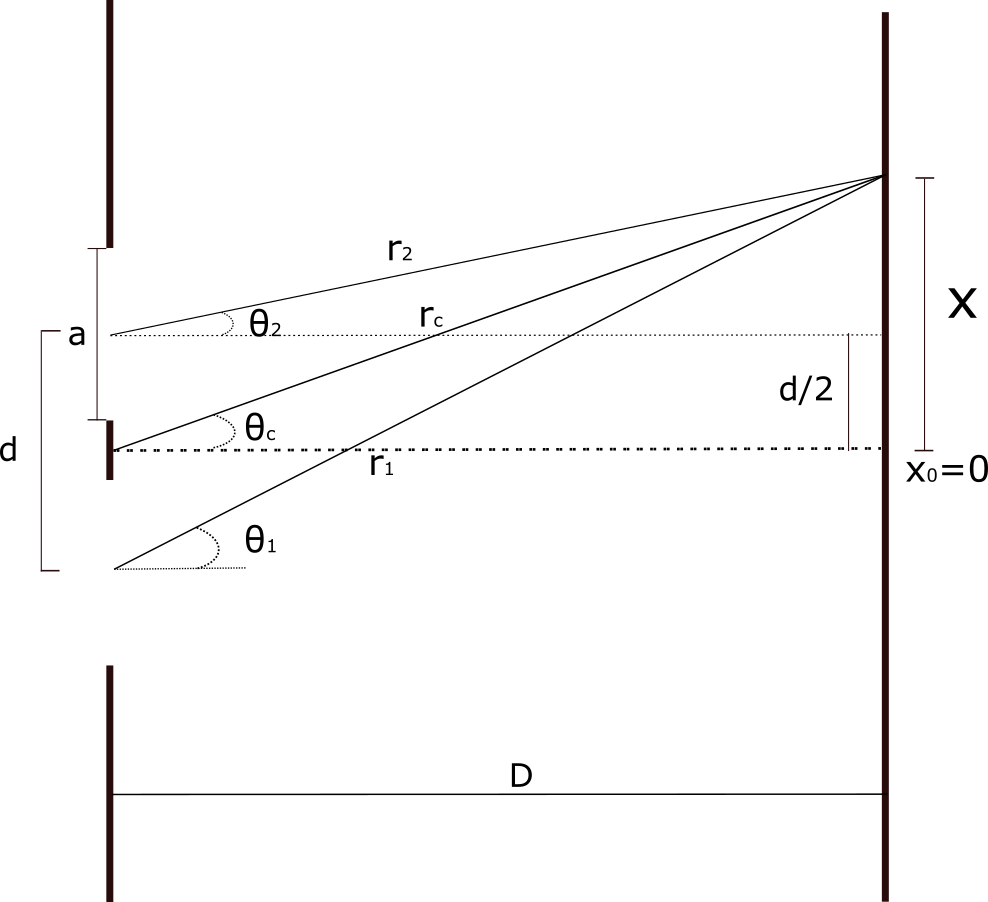}
\caption{Description of geometric double slit diagram with no small angle assumption.
}
\label{fig:geometric}
\end{figure}

\begin{figure}
\includegraphics[scale=.65]{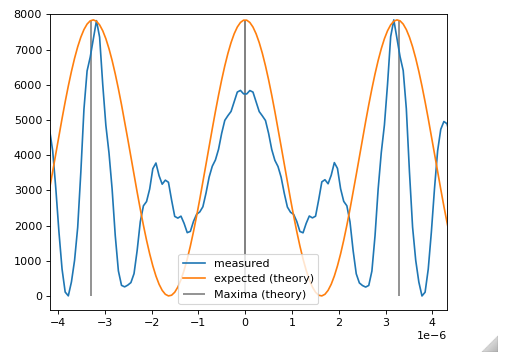}
\caption{Integrated Poynting magnitude recorded at the final time step. This recording was taken before the far-field was reached, thus the magnitude of the central interference band is not as large as the second band. Note the lower-order peaks which are undesirable for an ideal double slit.
}
\label{fig:integrated}
\end{figure}

In FIG.~\ref{fig:integrated}, we instead compare the integrated FieldProbe line's Poynting vector magnitude to expected values calculated analytically. In order to obtain accurate data for a small simulation, we calculated the expected diffraction pattern without making assumptions. Refer to FIG.~\ref{fig:geometric} for initially defined geometry.

We know from wave optics that the locations in the diffraction pattern with constructive interference are the locations where the difference between the distance the waves travel is an integer multiple of the wavelength.

\begin{subequations}
\label{eq:gamma}
\begin{equation}
    m\lambda = \delta = r_{2} - r_{1}\label{gameq:1}
\end{equation}
\begin{equation}
    \delta = D\left[\frac{1}{\cos(\theta_{2})} - \frac{1}{\cos(\theta_{1})}\right].\label{gameq:2}
\end{equation}
\end{subequations}
We now unify $\theta_{2}$ and $\theta_{1}$
\begin{subequations}
\label{eq:theta}
\begin{equation}
    \theta_{2,1} = \arctan\left[\frac{x\pm d/2}{D}\right]\label{thetaeq:1}
\end{equation}
\begin{equation}
    \theta_{C} = \arctan\left[\frac{x}{D}\right].\label{thetaeq:2}
\end{equation}
\end{subequations}

By solving $\theta_{C}$ for $x$ and substituting into Eq.~\eqref{gameq:2}

\begin{eqnarray}
    m\lambda = D\left[\frac{1}{\cos( \arctan (\tan(\theta_{c})+d/2D))}-\right.\nonumber\\
    \left.\frac{1}{\cos(\arctan(\tan(\theta_{c})-d/2D))}\right].    \label{mlam}
\end{eqnarray}
Now by solving for $\theta_{C}$
\begin{equation}
    \label{thetacent}
    \theta_\mathrm{c-bright} = \arctan\left[\frac{m\lambda\sqrt{-4D^2-d^2+m^2\lambda^2}}{D\sqrt{-4d^2+4m^2\lambda^2}}\right].
\end{equation}

Using this equation, the target locations of maximum constructive interference are then calculated trigonometrically using
\begin{equation}
    \label{trig}
    X_\mathrm{bright} = D\tan(\theta_{C})
\end{equation}

Similarly, the interference pattern with respect to X can be calculated by using phasors. We can find the phase difference ($\phi$) in the slits via
\begin{equation}
    \label{phi}
    \phi = 2\pi \frac{d\sin\theta}{\lambda}
\end{equation}
and because the intensity is a factor of $E^2$, when we double the E (for constructive interference), a scaling factor of 4 emerges.
\begin{equation}
    \label{intensebase}
    I=4I_{0}\cos^2\left(\frac{\phi}{2}\right)
\end{equation}
by substituting, we find\cite{2020Double}
\begin{subequations}
\begin{equation}
    \label{int1}
    I=4I_{0}\cos^2\left(\frac{d \pi \sin \theta}{\lambda}\right)
\end{equation}
\begin{equation}
    \label{int2}
    I=4I_{0}\cos^2\left(\frac{d \pi}{\lambda}\frac{x}{D}\right)
\end{equation}
\end{subequations}

Eq.~\eqref{int2} yields the interference pattern. From phys.libretexts\cite{2020Double} we can also find an equation for the intensity diffraction caused by the split. See Eq.~\eqref{eq:beta}.
\begin{subequations}
\begin{equation}
    I = I_0\left(\frac{\sin\beta}{\beta}\right)^2
    \label{eq:beta}
\end{equation}
\begin{equation}
    \beta = \frac{\phi}{2} = \frac{a\pi\sin\theta}{\lambda}
\end{equation}
\end{subequations}

\begin{figure}
\includegraphics[scale=.4]{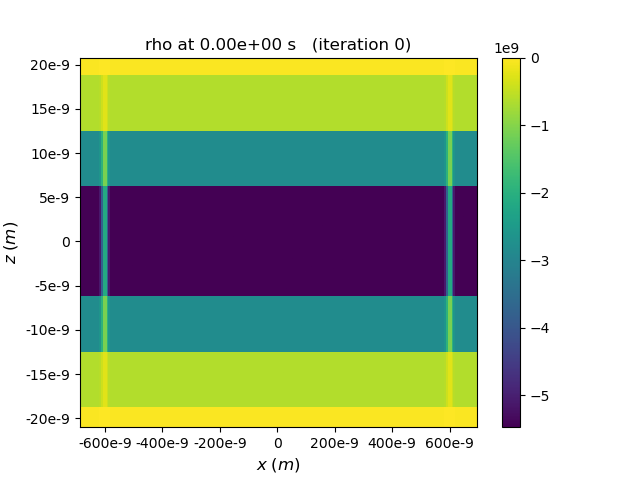}
\caption{The particle density is seen highly magnified. The double slits are the vertical breaks in the density. Note that there is still a nonzero density of electrons within the slits. A \textit{fuzziness} can be seen interacting with our plane wave, resulting in skewed data.
}\label{fig:densityfail}
\end{figure}

Plotting the analytically expected maxima locations and intensity curves along with the measured values produces FIG. (\ref{fig:integrated}). As can be seen, the peaks and curvature do not closely match expected values. There is an intensity drop in the central maxima that occurs because this measurement was taken on the outer edge of the near field. To obtain proper intensity in the central maxima, the simulation would need to extend into the far field. However, several small, second order local maxima also appear. From analyzing openPMD data, these appear to be the result of the simulation setup using a particle density for the double slit. By investigating further into the particle density, we find 2 inherent issues with using this form of particle density to act as our slit. The first, as can be seen in FIG.~\ref{fig:openpmd_double}, is that a large portion of the plane wave permeates the boundary, resulting in a large offset for outputed data. This could be counter-acted with a higher-density plasma, but this in turn requires a finer resolution of its plasma frequency to compute. The second, as can be seen in FIG.~\ref{fig:densityfail}, is that because of the algorithm used by WarpX to place particles, and because these particles have an inherent geometry, a \textit{fuzziness} appears at the slit locations. The desired result is to instead have a clean double slit. To solve this problem, we replaced the particle density with an embedded metal boundary occupying the same dimensions, in which the electromagnetic fields are analytically set to zero.

\begin{figure}
\includegraphics[scale=.5]{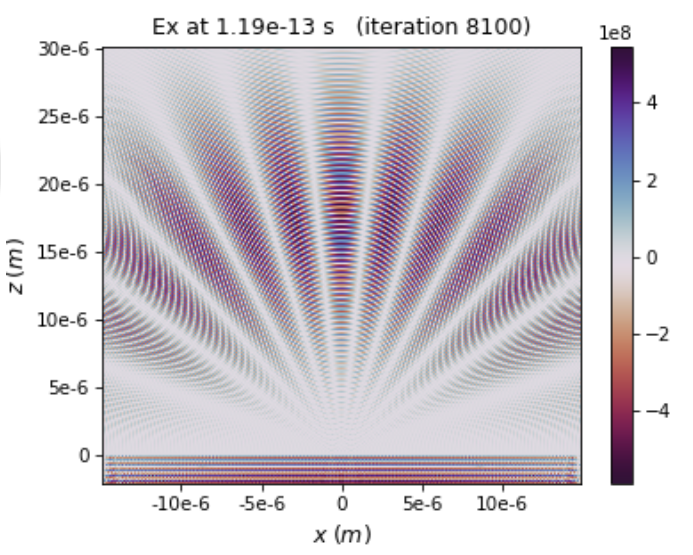}
\caption{Using the embedded boundary, a more idealized double slit is achieved. Note the clear distinction between constructive and destructive interference patterns.
}\label{fig:Perfect}
\end{figure}

\begin{figure}
\includegraphics[scale=.40]{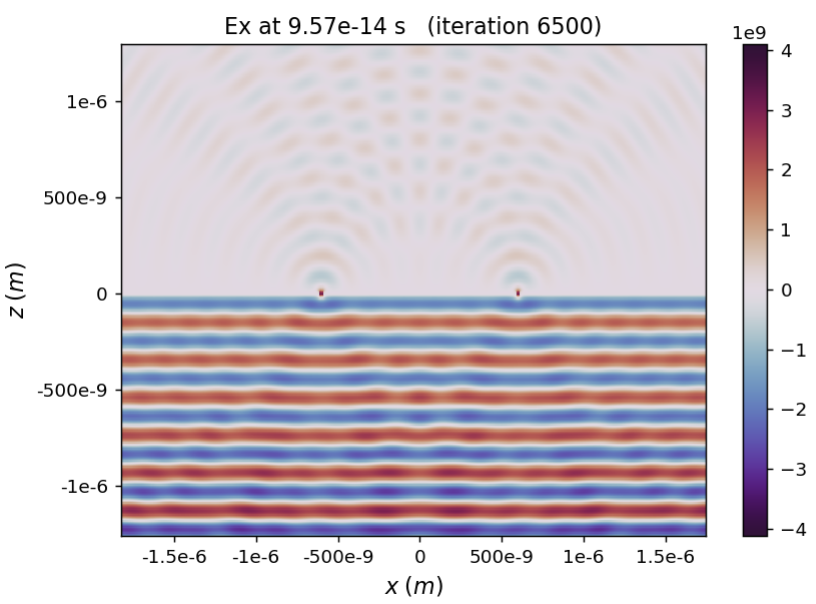}
\caption{A close up view of the double slit using the embedded boundary. Note that there is no plane wave permeating through the metal boundary. A clean diffraction pattern is visible.
}\label{fig:SlitClose}
\end{figure}

\begin{figure}
\includegraphics[scale=.38]{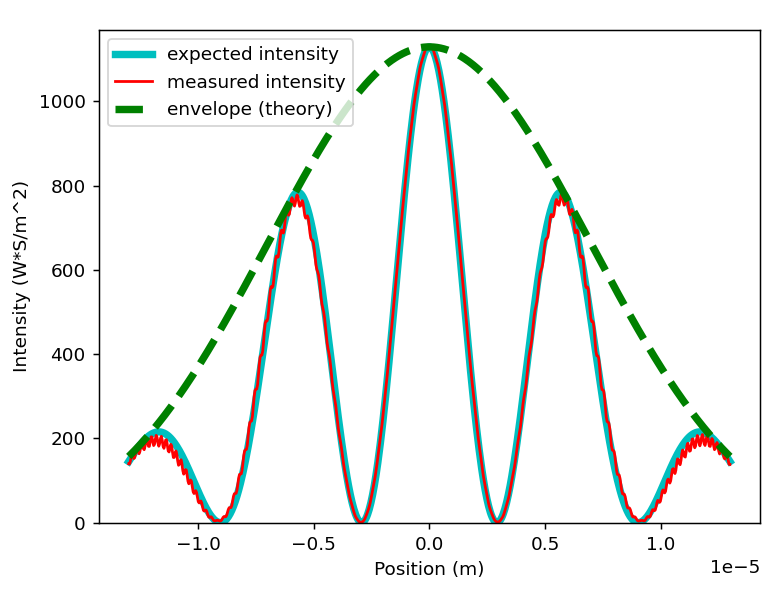}
\caption{A measurement of the measured intensity versus the analytically calculated expectation. The red line (which traces the experimental results in blue) is the product of the interference and diffraction patterns.\cite{2020Double}
}\label{fig:Final}
\end{figure}

Figures~\ref{fig:Perfect}, \ref{fig:SlitClose}, and \ref{fig:Final} were created using a larger simulation domain ($42.5\,\upmu$m in the Z direction and $40\, \upmu$m in the X direction. By allowing the simulation to run over longer distances in the Z direction, the waves approach the analytical solutions (which assume measurements in the far field). By utilizing the embedded boundary, we managed to cut out the noise created via waves permeating through the slit walls (see FIG.~\ref{fig:SlitClose}). Additionally, because this boundary is not defined by particle geometries, but rather precise locations, we managed to increase the clarity of the slit while decreasing the sheet width to the size of 2 cells (12.5\,nm). This small size (relative to the wavelength and simulation domain) allowed us to make the simplifying assumption that the two waves are emanating from single points rather than channels.

By taking the product of Eq.~\eqref{int2}) and Eq.~\eqref{eq:beta}, we can calculate a function for the expected intensity as a function of x. By plotting the interference pattern, intensity diffraction, the expected intensity function, and the experimental results, we are able to produce FIG.~\ref{fig:Final}. This figure shows a near perfect match between analytically expected and experimentally observed intensity. This figure, as well as FIG.~\ref{fig:line_out} demonstrate the effectiveness and accuracy of the \textit{FieldProbe} diagnostic.
\section{Conclusion}
The FieldProbe diagnostic is prepared to be utilized in WarpX to calculate electromagnetic flux as well as to find instantaneous field values throughout the simulation. The field probe can be used in tandem with any other diagnostic and can be implemented with a single point detector, line detector, or plane detector. Multiple FieldProbe diagnostics can be active at the same time, with independent spatial and temporal fidelity. The probe has been shown to be accurate when compared to analytically calculated values as well as values found via pre-existing diagnostics.

For future iterations, the field probe can be expanded to spectrally filter certain wavelengths in an attempt to reduce noise or model wavelength-filters in experiments, to travel with a wave as it is moving, and to include Lorentz transformations in the calculations of boosted-frame simulations.
\section{Acknowledgements}
Special thanks goes out to Dr. Axel Huebl for his hands-on mentoring during the development of this project.
I would like to acknowledge Berkeley Lab’s Accelerator Modeling program and its associated members for their advice and direction towards developmental resources as well as all WarpX contributors.
I would also like to thank Dr. Pickett, Dr. Jaikumar, and Dr. Kwon of California State University, Long Beach for their continued academic support.
This work was prepared in partial fulfillment of the requirements of the Berkeley Lab Undergraduate Research (BLUR) Program, managed by Workforce Development \& Education at Berkeley Lab. 
This project was supported by the Exascale Computing Project (17-SC-20-SC), a collaborative effort of two U.S. Department of Energy organizations (Office of Science and the National Nuclear Security Administration) responsible for the planning and preparation of a capable exascale ecosystem, including software, applications, hardware, advanced system engineering, and early testbed platforms, in support of the nation's exascale computing imperative.
This work was performed in part under the auspices of the U.S. Department of Energy by Lawrence Berkeley National Laboratory under Contract DE-AC02-05CH11231.
\section{References}
\bibliography{main}
\end{document}